\documentclass[12pt]{article}

\usepackage{epsfig}

\begin{document}

\begin{center}
{\Large\bf The Higgs Particle: what is it, and why} \vspace*{3mm} \\
{\Large\bf did it lead to a Nobel Prize in Physics?}
\end{center} 

\vspace*{1mm}
\begin{center}
Wolfgang Bietenholz \\
Instituto de Ciencias Nucleares \\
Universidad Nacional Aut\'onoma de M\'exico \\
A.P.\ 70-543, C.P.\ 04510 Distrito Federal, Mexico
\end{center}

\vspace*{4mm}

{\em Back in 1964, the theoretical physicists Fran\c{c}ois Englert
and Robert Brout, as well as Peter Higgs, suggested
an explanation for the fact that most elementary particles
--- such as the electron --- have a mass. This scenario predicted
a new particle, which has been observed experimentally only just now
at CERN (the European Organization for Nuclear Research). 
This discovery led to the Physics Nobel Prize 2013.
Here we sketch in simple terms the concept of the 
Higgs mechanism, and its importance in particle physics.} \\

To the best of our knowledge, the world consists of only
very few types of elementary particles, the smallest
entities of matter, which are indivisible.
They are described successfully by the Standard Model (SM)
of particle physics, a great scientific 
achievement of the 20th century. All phenomena observed so far
with elementary particles are compatible with the SM, 
which made a large number of correct predictions.

There is {\em one} particle that the SM needs in order
to work, which has been observed only very recently: 
the famous {\em Higgs particle}. After intensive
and careful work, the collaborations ATLAS and CMS, working 
at the Large Hadron Collider at CERN near Geneva (Switzerland),
reported in December 2011 first hints of its observation.
These hints were further substantiated in 2012, and the discovery
of the Higgs particle is now generally accepted. Therefore Englert 
and Higgs have been awarded the 2013 Nobel Prize in Physics for
their correct prediction (Brout passed away in 2011). 
\begin{figure}[h!]
\begin{center}
\includegraphics[angle=0,width=.62\linewidth]{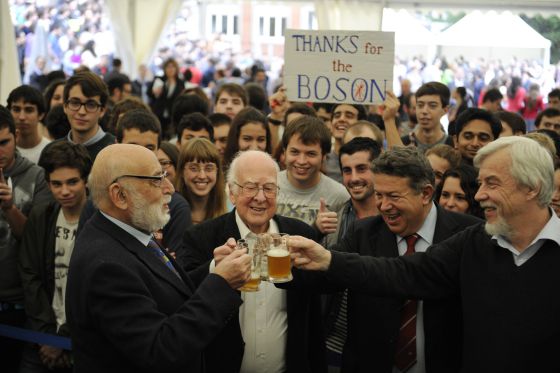}
\end{center}
\caption{\it{Fran\c{c}ois Englert (on the left), Peter Higgs (center-left)
and CERN director Rolf-Dieter Heuer (on the right) celebrating the
discovery of the Higgs boson, and the consequential Nobel Prize.
The existence of this particle had been predicted in 1964.
It was finally confirmed in 2011/2 at CERN.}}
\end{figure}
\

Hence the Higgs particle is now in the focus
of interest in physics, and also in popular science. Unfortunately,
the latter often denotes it as the ``particle of god'', which sounds
spectacular, but which does not make any sense whatsoever.
If one assumes the creation of the Universe by some kind of
god, then {\em all} particles are ``particles
of god'', and otherwise none is linked to theology,
but there is no way to assign this r\^{o}le specifically
to the Higgs particle. Here we hope to disseminate a better
view what this myth-enshrouded particle is about. 
\begin{figure}[h!]
\begin{center}
\includegraphics[angle=0,width=.62\linewidth]{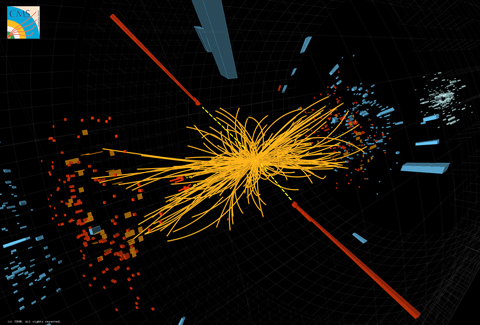}
\end{center}
\caption{\it{Tracks of elementary particles detected by the CMS 
Experiment at CERN in Geneva. The picture shows an event, which could 
give a hint about the Higgs particle, with two high energy photons
(red bars) and other particles (yellow lines), generated in
a powerful proton--proton collision.}}
\end{figure}
\\

The SM is formulated as a {\em Quantum Field Theory.} In physics, 
{\em fields} are abstract functions of space and time,
{\it i.e.}\ on each point and at any time some field value
is introduced.\footnote{Here we refer to the ``functional
integral formulation'' of Quantum Field Theory.
Alternatively, the ``canonical formulation'' deals with
operator valued fields, but the resulting physics is the same.}
This could be the temperature or the pressure in each point of
a hall during one hour, or in an ocean during one year.
A field may also have several components, which can be of
a more abstract kind than real numbers.\footnote{Field
values can also be given by vectors, tensors or matrices (representing
group elements), and their components could be complex numbers, 
or anti-commuting ``Grassmann numbers''.} If we assign a specific
field value to each space-time point under consideration, we
obtain one {\em configuration.} 

The occurrence of elementary particles is described 
by various types of fields. Their properties and dynamics are
characterized by a function of the field configurations involved, 
known as the {\em action.}
Classical field theory only considers one specific configuration
of each field, the one that minimizes the action. For instance,
if one performs this minimization for the electromagnetic fields,
one obtains the Maxwell equations. 

Quantum Field Theory, however, keeps track of the sum over all 
possible configurations. The one with minimal action or
energy\footnote{For simplicity, our discussion treats
the {\em action} and {\em energy} as equivalent
properties of a given field configuration. A transition
to ``imaginary time'' 
is --- for equilibrium states --- a mathematically allowed
transformation, which justifies this identification.}
--- it could be the zero fields ---
corresponds to the absence of particles, the {\em vacuum.} 
The additional energy that it takes for a 
fluctuation around this vacuum to generate just one particle is 
the {\em mass} of this particle.

Various fields --- and the associated particles --- may feel each other, 
{\it i.e.}\ they can interact, if the action contains a product
of distinct fields that occur at the same point. In our understanding 
of the emergent complicated systems, {\em symmetries}
play a key r\^{o}le. A symmetry is a group of transformations of 
the fields, which do not alter the action, so they are in general
not observable.
We distinguish {\em global} and {\em local} symmetries. A {\em global}
symmetry allows us to change a field in the same way all over
space and time --- like an Aerobic session where many people move 
simultaneously in the same manner.
{\em Local} symmetries are even more stringent:
here the field can be changed in each space-time point in a
different way, and still the action remains invariant.
That appears like a chaotic Aerobic session, where everyone 
moves as he or she likes.

If one requires such a local symmetry to hold, a huge a 
number of field transformations are allowed, and it is a delicate
challenge to maintain invariance under all of them. For the
actions that one usually starts with, this is not the case 
--- they do change under
most local field transformations. However, one can repair the
invariance by introducing additional fields, which transform
exactly such that these changes are compensated. These are
the {\em gauge fields}, which transmit an interaction
between the ``matter fields'' that we had before. Moreover
they represent own kinds of particles, such as the 
{\em photon} (the particle of light).
In fact, the dominant interactions among the
SM particles are transmitted by a set of gauge fields.
This only works if the local symmetry is preserved exactly.

When this concept was developed, people noticed its virtues,
but also a severe problem: the requirement of a
local symmetry does not allow us to include any term in the
action, which would simply specify some energy that it costs
to ``switch on the field'', {\it i.e.}\ to deviate from
a zero configuration, and therefore to represent a {\em particle
mass} in its simplest form. Still we know
that particles like the electron do have a mass. This was
the puzzle that physics was confronted with in the 1950s,
and which was later overcome by the famous Higgs mechanics.

The idea of this mechanism is that one does not necessarily
need to refer to the zero field configuration. Instead 
one couples for instance the electron field to a new
{\em Higgs field}, which is endowed with a self-interaction 
such that it takes its energy minimum for non-zero configurations. 
Then fluctuations away from this minimum
require some energy, specifying
a particle mass, while fully preserving the local symmetry.
Now there is a whole set of non-zero field configurations
corresponding to the minimal energy. The configurations in
this set are related by local symmetry transformations,
so physics is indeed invariant, and they all correspond to
the same vacuum.\\

To provide an intuitive picture, we refer to a historic
{\em Gedankenexperiment} (a ``thought experiment''), 
known as ``Buridan's donkey''. {\em Jean Buridan} was a 
French scholastic philosopher of the 14th century,
who was interested in logic, mechanics, optics, and in the existence 
and meaning of a ``free will''.

\noindent
\rule[-0.1cm]{13.5cm}{0.01cm}

\vspace*{1mm}
{\small{Regarding Buridan's biography, we know that
he that was born in France around 1295, he studied 
philosophy at the University of Paris, where he was subsequently
appointed professor in the Faculty of Arts, and also rector for two 
years. Around 1340 he condemned the views of his teacher and mentor 
William of Ockham, which has been interpreted as the dawn of
religious skepticism and the scientific revolution.
In the 15th century, Ockham's partisans placed Buridan's works 
on the Index of Forbidden Books.
\begin{figure}[h!]
\vspace*{-1mm}
\begin{center}
\includegraphics[angle=0,width=.4\linewidth]{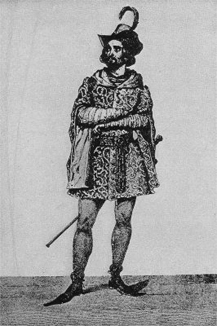}
\end{center}
\vspace*{-5mm}
\caption{\it{Jean Buridan, French philosopher and scientist of the 14th 
century, best known for the Gedankenexperiment with a hungry donkey.}}
\vspace*{-1mm}
\end{figure}

Beyond that, Buridan's life is even more myth-enshrouded than 
the Higgs particle: according to some sources, he was forced to 
flee from France, spent time in Germany and also founded 
the University of Vienna in 1356 (or at least attended its foundation). 
Other records describe him as a charismatic and
glamorous figure with numerous amorous 
affairs, which even involved the French Queen Jeanne de Navarre.
Therefore King Philippe V (supposedly) sentenced him to be thrown in 
a sack into the Seine River, but he was saved by one of his students.
Still another legend claims that he violently hit Pope Clement VI 
over the head with a shoe, trying to gain the affection of 
a German shoemaker's wife. 

Buridan died around 1358, possibly
as a victim to the Black Plague.}}
\vspace*{-3mm} \\
\rule[-0.1cm]{13.5cm}{0.01cm} \\

As a general background, even before classical mechanics was
worked out mathematically,\footnote{Buridan himself worked
on a theory of ``impetus'', which is similar
to our modern term ``momentum''.} scholars often had an entirely
deterministic view of the world. In fact, mechanics seems 
to suggest that the course of any future evolution
is strictly determined by the present state of the Universe, 
given by the current positions and velocities of all objects.
Then the future should
follow an inevitable pattern, like a huge machine proceeding
step by step in a fully predictable way. Without knowledge
about quantum physics (and discarding sudden jumps against the
Laws of Nature), it is not obvious to find an objection against
this picture.

However, its strict application would even capture mental
processes in the brains of animals and human beings. This
conclusion appeared confusing, since it implies that our
``free will'' is a plain illusion. In the context of this
discussion, {\em Buridan's donkey} was invented as a fictitious, extreme 
example.\footnote{According to some sources, it was actually
invented by Buridan's opponents to ridicule the deterministic
point of view, by {\it reductio ad absurdum}.}
One imagines a hungry donkey, to whom one offers two
piles of hay. However, they are fully identical and placed
at exactly the same distance to its left and to its right.
Hence the donkey has to make a decision for one direction
in order to be able to eat. If nothing favors one of the
piles, and its mind is fully deterministic, the poor
donkey will stay in the middle and finally starve to death,
although its salvation is so close. Taking a sudden decision 
corresponds to a process, which is denoted
as ``Spontaneous Symmetry Breaking'' in Quantum Field Theory.
Upon arrival at one of the piles, the donkey does not perceive
the left/right symmetry (or ``parity'') anymore.

\begin{figure}[h!]
\begin{center}
\includegraphics[angle=0,width=.63\linewidth]{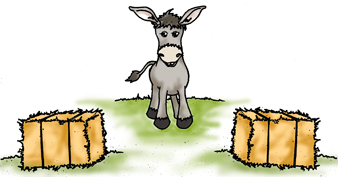}
\end{center}
\caption{\it{Buridan's donkey between two piles of hay, faced with the
dilemma if it should walk to the left or to the right in order
to eat. Taking a decision corresponds to the process
of Spontaneous Symmetry Breaking.}}
\end{figure}

To translate this setting into the Higgs mechanism, we should
better talk about a thirsty donkey in the center of a circular
water ditch. Now there is a continuous set of favored positions 
--- corresponding to the energy minimum --- anywhere next to 
the ditch. If we sum over all possible positions 
(all possible ``donkey field configurations''), these favored 
positions provide the statistically dominant contributions. So if 
we evaluate the expectation value of its water supply, we conclude
that the quantum donkey is better off than its classical cousin:
it is able to drink. 
This reveals the importance of Quantum Field Theory\footnote{Even in 
Quantum Mechanics --- the theory that preceded Quantum Field Theory in 
the first half of the 20th century --- the phenomenon of Spontaneous 
Symmetry Breaking does not occur.}: in fact, it can be {\em live saving!}
\begin{figure}[h!]
\begin{center}
\includegraphics[angle=0,width=.7\linewidth]{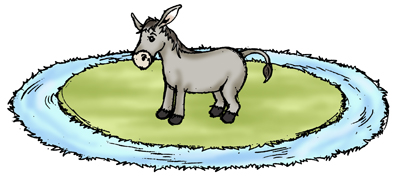}
\end{center}
\caption{\it{A modified donkey, now thirsty and surrounded by 
a water ditch. There is an infinite number of directions 
where it could go in order to drink. Its preferred positions are
displaced from the starting point (at zero), next to the water. 
If it arrives there somehow, a motion along the ditch corresponds 
to a massless particle, known as a Nambu-Goldstone boson.}}
\end{figure}

Once the donkey has attained the water, it can freely move
along the ditch and keep on drinking. This kind of motion
keeps the energy at its minimum. The corresponding
fluctuation of the donkey field configuration does not
cost any energy, hence it corresponds a massless particle,
known as a {\em Nambu-Goldstone boson.} 

If we now couple the donkey field --- which takes the r\^{o}le
of the Higgs field --- to other fields, such as the one of
the electron, the shift of the favored position away from the
center ({\it i.e.}\ away from the zero configuration) 
yields the electron mass.

With gauge fields included, all positions at the ditch
(with the energy minimum) are physically identical, since they
are now related by local symmetry transformations. Thus a walk
along the ditch is not a real motion anymore, and the
Nambu-Goldstone bosons disappear again. Instead some of the gauge
particles pick up a mass, in a subtle indirect way, which
fully preserves the local symmetry.\footnote{The literature
often calls this process a ``spontaneous gauge symmetry breaking'',
although any local symmetry (or gauge symmetry) transformation 
preserves the physical state, which solely describes the
donkey's proximity to the water. Therefore, strictly speaking
the gauge symmetry does {\em not} break.

In the SM this process provides masses for the three gauge bosons
of the weak interactions ($W^{\pm}$, $Z^{0}$), and for all the 
fermions (quarks and leptons), but not for the photon and 
the gluons, which transmit the electromagnetic and the strong
interaction, respectively.}

This happens for a suitable system a low temperature (in an infinite 
volume). The situation at high temperature could be sketched as
pouring a lot of water into the area, hence the donkey does
not need to move in order to drink. Then the
procedure works with respect to its zero position, and no
symmetry breaking occurs.\\

\begin{figure}[h!]
\begin{center}
\includegraphics[angle=0,width=.55\linewidth]{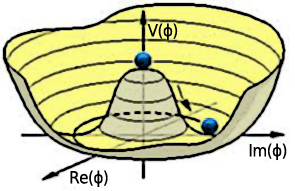}
\end{center}
\caption{\it{A potential $V ( \phi)$ for a complex scalar field $\phi$,
which takes its minimal energy for non-zero field configurations
$| \phi| > 0$. There is a continuous set of minima. Fluctuations
within this set correspond to a Nambu-Goldstone boson. Once the field 
is gauged, all minima are physically identical, the Nambu-Goldstone 
boson disappears, but the gauge field picks up a mass.}}
\end{figure}

Let us repeat that any deviation away from the vacuum 
state costs energy, and here we capture the masses, 
for the electron and for other particles,
without breaking the sacred principle of local symmetry. 
Several physicist noticed this property
in the early 1960s. A corresponding mechanism
was known in solid state physics, and applied also to
particle physics by Fran\c{c}ois Englert and Robert Brout
(in Bruxelles, Belgium), and independently by Peter Higgs.
In particular, it was Peter Higgs --- at that time a young lecturer 
at the University of Edinburgh (Scotland) --- who (encouraged by 
Yoichiro Nambu) pointed out that this mechanism, when applied to 
particle physics, brings about a new particle, which should be observable.
Its mass, however, could not be predicted,\footnote{The ``triviality'' 
of the Higgs model (see below) implies at least an upper limit 
for the theoretically possible mass of the Higgs particle.}
and the emerging mass of the other SM particles neither. 
This may be considered as a short-coming of
the SM: it contains a number of free parameters (about 26,
neutrino masses included),
which one would like to reduce. On the other hand, for describing
practically the whole Universe, this number is {\em not} alarming.
For comparison, many fashionable theories beyond the SM (like 
``supersymmetry'') do not only lack any observational support, 
but they introduce in addition an avalanche of further free parameters.
\begin{figure}[h!]
\begin{center}
\includegraphics[angle=0,width=.55\linewidth]{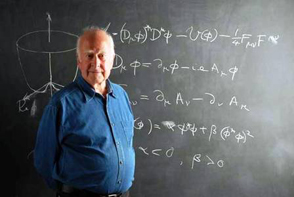}
\end{center}
\caption{\it{Peter Higgs, explaining the theory that predicts the
famous particle named after him.}}
\end{figure}

Now the observation of the Higgs particle has been confirmed, 
so we can feel proud of a very well-established and elegant theory 
that describes all the elementary particles that we know of.
So are we then done, and physicist will end up unemployed?

Not really, even the great SM has its short-comings, 
that we still have to work on:

\begin{itemize}

\item It does not capture all interactions:
the most obvious one in contemporary life, {\em gravity,} is not
included. Intensive attempts (over several decades) 
to incorporate it have failed. Gravity is 
described successfully by a different theoretical
framework, Einstein's Theory of General Relativity, which 
seems simply incompatible with Quantum Field Theory.
While this is an outstanding challenge, for practically
all issues in particle physics it can be ignored, since
gravitational effects are usually negligible in the microscopic 
context (an exception was the very early Universe).

\item We have nowadays indirect but clear evidence of further 
ingredients to the Universe, denoted as {\em Dark Matter} and 
{\em Dark Energy.}
Their nature is mysterious, and the SM cannot capture them ---
another tremendous challenge to work on.

\item Even with known matter --- consisting of the SM
particles --- {\em complex structures,} as they occur for instance
in biology, are not simply understood based on the SM
as the fundamental theory.
Here a deep understanding of the collective behavior of many 
particles has to be supplemented, which has been accomplished 
only in part.

\item Finally the SM has an intrinsic reason for being incomplete. 
A na\"{\i}ve treatment of Quantum Field Theories yields infinities in
quantities that we want to compute. They diverge when we take
field fluctuations at all energy scales into account.
We can render them finite by introducing an energy cutoff,
which should be done in a subtle way, preserving again the
local symmetries. In some cases we can later --- in the final
result for physical observables --- remove this cutoff by sending 
it to infinity. However, in case of the Higgs sector of the
SM this does not work: it would lead to a decoupling of the Higgs
field from all other fields, and therefore again to vanishing particle
masses; that property is known as {\em triviality.} 

So we have to live with such an energy cutoff, which is acceptable as long 
as it is far above the Higgs particle mass, but not high enough to render 
the Higgs field ``trivial'', {\it i.e.}\ free of interactions.
This implies that the validity of the SM is limited
to a certain energy range --- at even higher energies it requires
the extension to a superior theory, that we do not know yet.
The specialized literature suggests candidate theories in abundance,
but so far none has been substantiated.

\end{itemize}

Nevertheless, even if we find one day corrections to the SM
(under extreme conditions), it will always remain the appropriate 
description of particle physics in the energy range, which is most 
relevant to us --- just like Newton's theory of gravity, or the 
continuous description of thermodynamic systems, 
remain highly useful, although they are not exact.

\vspace*{2.2cm}

\noindent
{\small I am indebted to Aline Guevara and Eduardo Serrano for 
their assistance with the illustrations.
A shorter Spanish version of this article has been
published --- together with Daniella Ayala Garc\'{\i}a ---
in the Bolet\'{\i}n de la Sociedad Mexicana de 
F\'{\i}sica, vol.\ 26 no.\ 3 (2012) pp.\ 161-166.}

\end{document}